\documentstyle[]{article}

\begin{document}

\title{Simplifying Tensor Polynomials with Indices}
\author{
 A. Balfag\' on\thanks{Institut Qu\'\i mic de Sarri\` a, Laboratori de F\'\i sica Matem\` atica,
Societat Catalana de F\'\i sica (I.E.C.)Universitat Ram\' on Llull e-mail: abalf@iqs.url.es} and X. Ja\' en.\thanks{Universitat Polit\` ecnica de Catalunya, Laboratori de F\'\i sica Matem\` atica,
Societat Catalana de F\'\i sica (I.E.C.)
e-mail: jaen@baldufa.upc.es}
}
\date{}
\maketitle
\begin{abstract}
We are presenting an algorithm capable of simplifying tensor polynomials with indices when the building tensors have index symmetry properties. These properties include simple symmetry, cyclicity and those due to the presence of partial and covariant derivatives.

We are also including some examples using the Riemann tensor as a paradigm.

The algorithm is part of a {\it Mathematica} package called {\it Tools of Tensor Calculus (TTC)}[web address: http://baldufa.upc.es/ttc]
\end{abstract}
\section{Introduction}
The two main languages commonly used in writing tensor calculus expressions are the intrinsic notation and the index notation. The intrinsic notation seems to be the preferred language for  making computer algorithms  work with symbolic tensor expressions.  The index notation, on the other hand, is extremely powerful in expressing and manipulating tensors. There are some simple expressions (in index notation) which are difficult to express in intrinsic notation for the sake of introducing new operators with new properties. This is the case such as expressing
the simple operation of raising an index like  (\ref{eq1})

\begin{equation}
T_{i\ k}^{\ j}=g^{j m}T_{i m k} \label{eq1}
\end{equation}
(summation over repeated index is understood).

Recently  some computer tools capable of  working with index notation at a symbolic level have been appeared\cite{Christensen,Ilyin,Balfagon}.

In the present study we are facing to the problem of tensor polynomial simplification using index notation.
In a recent article\cite{Balfagon} we dealt with this problem, restricted to the case of simple index symmetry properties of the building tensors like 
\begin{equation}
T_{ i_{\sigma (1)}... i_{\sigma (n)}}+a T_{i_{1} ...i_{n}}=0\label{mono}
\end{equation}
 $a$ being a scalar and $\sigma$ a permutation of $\{ 1,...,n\}$.
 (\ref{mono}) includes symmetry and antisymmetry of any sequence of indices, and pairsymmetry. 
We will refer to these kind of properties as {\it homogeneous monoterm}.
The algorithm proposed in \cite{Balfagon} does not use any kind of library, neither temporal nor permanent and for that reason it does not require a lot of computer memory. 
V.A. Ilyn and A.P.Kryukov \cite{Ilyin} propose a method capable of working with index symmetry properties  which includes cyclic type properties like

\begin{equation}
T_{i j k}+T_{k i j}+T_{j k i}=0
\end{equation}
These were included by  the authors in a set called {\it multiterm like properties}.
Although the problem was satisfactorly solved theoretically, the computer time and the amount of computer memory needed to make elementary simplifications was large enough to consider that his study may have some practical and fundamental limitations ( since an expression with 11 index needs about 2500Mb). 

One of the problems encountered by V.A. Ilyn and A.P.Kryukov is related to the dummy indices. They consider the possibility of renaming dummy indices as new relations to add to  the set of relations generated using the true index properties of building tensors. So in the monomial as  $S_{j k} T^{k j i}$ as well as the properties due to the fact that $S$ is a symmetric tensor they add the ones deduced from the fact that the indices $j, k$ are dummy, that is to say
 \begin{equation}
S_{j k} T^{k j i}-S_{k j} T^{j k i}=0
\end{equation}
The number of this kind of relations increase factorially with the number of dummy indices, so the algorithm needs to work very hard for monomials of interest.  

The algorithm that we present here, written in {\it Mathematica}\cite{Mathematica} language, is a part of the package {\it Tools of Tensor Calculus (TTC)}\cite{Castellvi1,Castellvi2,Balfagon,TTC}. It can work with monomials whose building tensors have both  monoterm and multiterm properties. In addition to dealing with  dummy indices it also handles  symmetry  (including  the partial derivatives) and antisymmetry of any sequence of indices, cyclic type properties and the Ricci properties due to the presence of two or more covariant derivatives.

Taking the Riemann tensor as a paradigm  our algorithm can handle monomials that have this tensor as a factor with the known properties:

\begin{eqnarray}
R_{i j k l}=-R_{j i k l}=-R_{i j l k}=R_{k l i j}\nonumber \\
R_{i j k l}+R_{i l j k}+R_{i k l j}=0\nonumber\\
R_{i j k l ,m ....,n}=R_{i j k l ,n ....,m}\nonumber \\
R_{i j k l ;m}+R_{i j m k ;l}+R_{i j l m ;k}=0\nonumber\\
R_{i j k l \underbrace{;m ....;n}_{k}}=a R_{i j k l ;\underbrace{;n ....;m}_{k}}+({\rm Riemann\  terms})_{\underbrace{;...}_{k-2}}\label{riemp}
\end{eqnarray}

We have added the possibility of introducing simplifying properties such as
\begin{equation}
R^{m}_{\ i m j}=R_{i j}\ \ ;\ \ R^{m}_{\ m}=R
\end{equation}
explaining how to do that. The above properties are very usefull because, although they may not be needed, from the theoretical point of view, they allow drastic reduction the amount of indices used by the algorithm , during calculation.

We have left  the inclusion of the properties derived from the dimension of the base space, for the Riemann tensor and for the generic case for a future study.

\section{The problem }
The problem can be stated as follows

Given a tensor monomial written in index notation ( dummy index  included) together with all index properties of the building tensors , find a unique canonical  equivalent polynomial. 

The canonical equivalent polynomial is derived from the complete set of monomial relations generated from the exhaustive application of all index properties of the tensor factors over the original monomial. The solving procedure is related to some ordering criterion of the set of monomials. 

\subsection*{Dummy indices}
Let  $M_{0}(I_{F},J_{D}^{\alpha}) $ be a given tensor monomial, with $F$ free indices $I_{F}=\{ i_{1},...,i_{F}\} $ and $D$ dummy indices $J_{D}^{\alpha}=\{ j_{\alpha(1)},...,j_{\alpha(D)}\} $ ,  $\alpha$ being any permutation of  $\{1,...,D\}$ 

Example:

\begin{equation}
M_{0}(\{ i,k,l\} ,\{ j,m,p\} )=S_{i j k} T^{j \  m}_{\ p} T^{p}_{\ l m}
\end{equation}

Due to the meaning of the dummy indices, if $I$ is the identity permutation the following identities hold

\begin{equation}
M_{0}(I_{F},J_{D}^{\alpha})=M_{0}(I_{F},J_{D}^{I})\ \ \ \forall \alpha
\end{equation}
This means that there will be many equivalent expressions for a given monomial. This is the first problem to be solved. 

In \cite{Balfagon} we looked at a method in which the monomial could be rewriten in a canonical form with respect to the dummy indices independently of the number of dummy indices  present in the monomial, without using many computer memory. That is to say, from all the  lists $J_{D}^{\alpha }$ only one is necessary,  $J_{D}$. 

Essentially the method consists of using an ordering criterion that can be applied to a monomial without prior consideration of all the monomials to be ordered.

We will call \verb+ PCanonic[0]+ the function implemented for the algorithm. We have
\begin{equation}
M_{0}(I_{F},J_{D}^{\alpha})
\buildrel{\verb+ PCanonic[0]+}\over{\Longrightarrow} M_{0}(I_{F},J_{D})
\end{equation}
In the following sections we will handle monomials in the form $M_{0}(I_{F},J_{D})$. In this way we will not mention the dummy indices problem anymore.

\subsection*{Tensor properties}

Of all tensor properties we will emphasize those that will be of importance within the algorithm. We are referring to the properties of single tensors of a given monomial. 
\begin{itemize}
\item {\bf Monoterm properties}: the result of their application over a given monomial is also a monomial. 

\item {\bf Multiterm properties}: the result of their application over a given monomial is a polynomial (more than one term). 

\item {\bf Homogeneous or non-ordered properties}: they are properties of only one tensor whose application over a given monomial does not guarantee a better (alphabetical) ordering result without considering other properties. 

Example

If we have $S_{i j}-S_{j i}=0$ or $T_{i j k}+T_{k i j}+T_{j k i}=0$  in principle we do not know how to use it to improve the ordering of the monomial  
$S_{i j k} T^{j \  p}_{\ m} T^{m}_{\ l p}$. We must take all the relations generated from the above-mentioned properties into account  in order to know which of the resulting monomials or polynomial is the best ordered.

\item {\bf Inhomogeneous or ordered properties }: These are properties relating more than one tensor,  which we know how to use to improve the order of a given monomial.

Example

The property $T^{m}_{\ m i}=A_{i}$ guarantees an improvement of the ordering of the monomial  $S_{i j k} T^{p }_{\ p m} T^{m}_{\ l j}$

These kind of properties can be used directly over the working monomial. 

Not all the property falls into these last two definitions. Clear counterexamples are the Ricci relation
\begin{equation}
	V_{i ;m ;n}-V_{i ;n ;m}=(V ({\rm Riemann\  terms})) \label{ricciid}
\end{equation}

From (\ref{ricciid}) we see that the right hand side terms can always be considered more ordered than the left hand side terms, although we do not know which of the left terms is the most ordered. In this case we must take into account all the possible substitutions of the homogeneous part in order to generate the properties of a given monomial.

The consideration of an ordered property can make  properties that in principle are not ordered become ordered or partially ordered properties. A clear example is the definition of the Ricci tensor in terms of the Riemann tensor $R^{m}_{\ i m j}=R_{i j}$  that turns the character of the Bianchi II identity of the Riemann tensor resulting 
\begin{equation}
R^{m}_{\ i j k;m}=R_{k i;j}-R_{i j;k}\label{riccibianchi}
\end{equation}
a partially ordered property, since we can consider that the terms containing the Ricci tensor are more ordered than the rest. If we want to include the definition of the Ricci tensor in the set of properties of the Riemann tensor we must include (\ref{riccibianchi}), as well as the properties belonging to the Ricci tensor itself. Although such a procedure can be difficult and unnecessary because we can work exclusively with the Riemann tensor, it is very convenient  to eliminate superfluous indices.
\end{itemize}

\subsection*{How to handle monoterm properties }

Given a monomial $M_{0}(I_{F},J_{D})$ the exhaustive application of the monoterm properties generate a set of relations of the type
\begin{eqnarray}
M_{0}(I_{F},J_{D})+ a_{o i}M_{i}(I_{F},J_{D})=0\nonumber \\ 
M_{i}(I_{F},J_{D})+ a_{i j}M_{j}(I_{F},J_{D})=0 \ i,j=0,1,...\label{sistema1}
\end{eqnarray}
where  $ a_{o i},a_{i j}$ are scalars.
This set of relations can always be solved in favour of the most ordered monomial. The result will be 
\begin{eqnarray}
M_{i}(I_{F},J_{D})= b_{i}M^{S}(I_{F},J_{D})\ \forall i=0,1,... 
\end{eqnarray}
$b_{i}$  being scalars and  $M^{S}(I_{F},J_{D})$ the most ordered monomial. The specific way to solve (\ref{sistema1})  will be explained in  section(3) .

We will call \verb+PCanonic[1]+ the function implemented by the algorithm
\begin{equation}
M_{i}(I_{F},J_{D})
\buildrel{\verb+ PCanonic[1]+}\over{\Longrightarrow} 
b_{i}M^{S}(I_{F},J_{D})\ \forall i=0,1,... 
\end{equation}

In the following section we will handle monomials in the form $M^{S}(I_{F},J_{D})$.

\subsection*{ How to handle multiterm properties }

Given a monomial $M_{0}^{S}(I_{F},J_{D})$ the exhaustive application of the multiterm properties generates a set of relations of the type
\begin{eqnarray}
M^{S}_{0}(I_{F},J_{D})+ \sum_{i} b_{o i}M^{S}_{i}(I_{F},J_{D})=0\nonumber\\ 
M^{S}_{i}(I_{F},J_{D})+ \sum_{j} b_{i j}M^{S}_{j}(I_{F},J_{D})=0\ \ \ i,j=0,1,... \label{sistema2}
\end{eqnarray}
where $b_{o i}, b_{i j}$ are scalars.

This set of relations can always be solved in favour of the set of the most ordered monomials that we will denote by $M^{S C}_{j}(I_{F},J_{D})$. In this way given any of the monomials $M_{i}^{S}(I_{F},J_{D})$, generated from  $M_{0}^{S}(I_{F},J_{D})$ by the multiterm properties, we have
\begin{equation}
M^{S}_{i}(I_{F},J_{D})= \sum_{j} c_{i j}M^{S C}_{j}(I_{F},J_{D})
\end{equation}
$c_{i j}$ being  scalars.

The specific way to solve (\ref{sistema2})  will be explained in  section(3) .

We will call \verb+ PCanonic[2]+ the function implemented by the algorithm
\begin{equation}
M^{S}_{i}(I_{F},J_{D})
\buildrel{\verb+ PCanonic[2]+}\over{\Longrightarrow} 
\sum_{j} c_{i j}M^{S C}_{j}(I_{F},J_{D})\ \forall i=0,1,... 
\end{equation}

\section{Algorithm description }

The algorithm for simplification of polynomial tensors in index notation, called \verb+SimplifyAllIndex+, has the following functions:

\subsection*{Functions acting on single tensors}
\begin{itemize}
\item \verb+InputTensor+:\\
\verb+ InputTensor[+{\it Tsymbol,basis,ranklist}\verb+]+ 
declares that the symbol ${\it TSymbol}$ will be a tensor in the basis {\it basis} and with rank and type  {\it ranklist}. 

Example
 
\verb+InputTensor[T,XX,{1,1,1,1}]+

\item \verb+InputSymmetries[+{\it Tsymbol}\verb+[+{\it index}\verb+]+,{\it symmetrispecifications}\verb+]+
declares that the tensor, previously introduced using \verb+InputTensor+,
{\it Tsymbol} has the symmetry properties of the indices, positioned as indicated by the argument {\it index}, according to the specifications {\it symmetriespecifications}. 

Example 1:

\verb+InputSymmetries[T[i,j,k,l],{i,k,l}]+ declares that  \verb+T+ is symmetric with respect the index  \verb+{i,k,l}+

Example 2:

here we declare the same properties as the Riemann tensor, see (\ref{riemp}), for the tensor \verb+T+ 

\begin{verbatim}
InputSymmetries[T[i,j,k,l],
	{{i,j}},{{k,l}},{{i,j},{k,l}},Cyclic[j,k,l]];

InputSymmetries[T[i,j,k,l,.;m],Cyclic[k,l,m]]
\end{verbatim} 

\item \verb+BasicRules[+{\it n}\verb+]+:
once the symmetries of indices have been declared the corresponding rules that implement these symmetries are stored in the list \verb+BasicRules[+{\it n}\verb+]+ . If the property is monoterm {\it n=1} and if it is multiterm {\it n=2}. The value {\it n}=0 is reserved for the case of ordered properties which are generally introduced explicitly.
\end{itemize}

\subsection*{Functions acting on monomials}
In the following we will assume that there are no ordered properties stored in \verb+BasicRules[0]+ as they are unnessencial. 

The main function is \verb+SimplifyAllIndex+:

\verb+Index[+{\it metricname},\verb+SimplifyAllIndex][+{\it polytens}\verb+]+ simplify, i.e. canonize, the polynomial {\it polytens} taking into account all the declared properties of the building tensor, as well as the dummy indices and the properties related to partial and covariant derivatives.   
\verb+SimplifyAllIndex+ is applied over each term of {\it polytens}. This function has, in turn, the following internal functions (We will suppose that the initial monomial is $M_{0}(I_{F},J_{D}^{\alpha})$):  

\begin{itemize}
\item \verb+PCanonic[0]+: first we canonize the monomial with respect to the dummy indices. This step is taken each time a monomial is the result of some kind of index manipulation. So  our initial monomial and all other generated monomials will have the form $M_{i}(I_{F},J_{D})$

\item \verb+LlistaMonomis[+{\it n}\verb+][+{\it count}\verb+[+{\it n}\verb+]]+: 
in these lists all the monomials considered before the present case are stored . 
{\it n}=1 for monoterm properties and {\it n}=2 for multiterm properties. {\it count}\verb+[+{\it n}\verb+]+ is a counter which says that the monomials therein are the {\it count}\verb+[+{\it n}\verb+]+-th studied monomials.

\item \verb+LlistaRules[+{\it n}\verb+][+{\it count}\verb+[+{\it n}\verb+]]+: 
in these lists the simplifying rules of the monomial in \verb+LlistaMonomis[+{\it n}\verb+][+{\it count}\verb+[+{\it n}\verb+]]+are stored. 

\item \verb+PCanonic[1]+
canonize a monomial with respect to the monoterm properties. First  look at the lists \verb+LlistaMonomis[1][+{\it count}\verb+[1]]+. If the working monomial is in one of these lists it applies the corresponding rule stored in  \verb+LlistaRules[1][+{\it count}\verb+[1]]+. If this is not the case it applies the rules stored in \verb+BasicRules[1]+ exhaustively  generating a list 
of new properties which will be solved by \verb+PFind+ giving as a result 
\verb+LlistaMonomis[1][+{\it count}\verb+[1]]+ and 
\verb+LlistaRules[1][+{\it count}\verb+[1]]+ and it applies these rules to the working monomial.

\item \verb+PCanonic[2]+
canonizes a monomial with respect to the multiterm properties once it has been canonized by \verb+PCanonic[1]+. First look at the lists \verb+LlistaMonomis[2][+{\it count}\verb+[2]]+. If the working monomial is in one of these lists it applies the corresponding rule stored in  \verb+LlistaRules[2][+{\it count}\verb+[2]]+. If this is not the case it applies the rules stored in \verb+BasicRules[2]+ exhaustively generating a list 
of new properties which will be solved by \verb+PFind+ giving as a result 
\verb+LlistaMonomis[2][+{\it count}\verb+[2]]+ and
\verb+LlistaRules[2][+{\it count}\verb+[2]]+ and applying these rules to the working monomial.

\item \verb+PFind+ is a simple tool that solves any system of linear equations with a fixed  ordering criterion.

Let $\{x_{1},x_{2},...,x_{M}\}$ be the space of ordered variables to be solved. Let 
\begin{equation}
\verb+eqlist+=\{c^{1}+a^{1 j} x_{j}=0,c^{2}+a^{2 j} x_{j}=0,...,c^{N}+a^{N j} x_{j}=0 \} \ \ \  \ j=1,...M
\end{equation}
be the list of $N$ equations with $M$ variables ordered first taking into account the variables and then the coefficients. That is to say, if in the $n$-th equation the most disordered variable is  $x_{m}$ then $x_{m+1}$ cannot be present in the $(n-1)$-th equation. 

\verb+PFind+ takes the initial list of equations and starts the following loop:

{\bf Step $1$:}
\verb+PFind+ takes the first equation in \verb+eqlist+. It is solved with respect to the most disordered variable. This result is stored in \verb+eqrule[1]+.
\verb+eqrule[1]+ is applied over \verb+eqlist+ and after eliminating zeros this result is assigned to \verb+eqlist+. Go to step 2

{\bf Step $n>1$:}
If \verb+eqlist+ is empty the loop is finished. On the contrary \verb+PFind+ takes the first equation in \verb+eqlist+. It is solved with respect to the most disordered variable. This result is stored in \verb+eqrule[+{\it n}\verb+]+.
\verb+eqrule[+{\it n}\verb+]+ is applied over \verb+eqrule[+{\it i}\verb+]+ $i=1,...,n-1$. \verb+eqrule[+{\it n}\verb+]+ is applied over \verb+eqlist+, the zeros are eliminated and the result assigned to \verb+eqlist+. Go to the step $n+1$.

\verb+eqlist+ will remain empty if the system is compatible. These must be the case for a set of compatible tensor properties. 
\end{itemize}

The computer time and memory of this algorithm have been controlled. The session presented in the appendix can be performed with no more than 2.6Mb. 

\section{Conclusion}

We have solved the problem of simplification of tensor polynomials in index notation with respect to a large set of properties for the building tensor satisfactorily. The computer time and memory needed are sufficiently reasonable so we think that the program can be of interest for practical purposes.

It is not our intention to say that TTC is a great program. Within {\it Mathematica} there are very good programs (See MathTensor \cite{MathTensor} ) Nevertheless we think that in the field of simplifying tensor polynomials we have reached a high level performance. Specifically, due the generic character of the algorithm presented (the only requirements is that the input polynomial follow the usual mathematical rules) makes that it can be useful in the kind of fields where tensors are a natural language. As is well know tensor calculus is power enough to cover large fields of mathematics, physics and engineering.

There is an important  set of properties not included in the program. These are the ones that appear when the dimension of the space is taken into account. 
The essential reason for not including these relations is that applying properties of the building tensor do not generate this kind of relations. They arise when the whole monomial is taken into account so we need some tool to generate the dimensional properties for a given monomial. We are leaving this new tool for a future study.

\section*{Acknowledgments}
X.J. would like to thank the Comisi\' on Asesora de Investigaci\' on Cient\' \i fica y T\' ecnica for partial financial support, under Contract No. PB96-0384. The authors would also like to thank the Laboratori de F\' \i sica Matem\` atica at the Societat Catalana de F\' \i sica for partial financial support.

\section*{Appendix: Mathematica Session. Demonstration of various properties relating Riemann monomials }

We present here an example of how the algorithm \verb+SimplifyAllIndex+ can be applied. We have chose the demonstration of well know polynomial properties of the Riemann tensor so the session will be also a short test for the algorithm.

The session is part of a {\it Mathematica} notebook. The inputs are indicated by 
\verb+In[]:=+   . The outputs have no prefix. The titles and comments are enclosed using the symbols \verb+(*+{\it comment}\verb+*)+. 

Since we use the Riemann tensor as example we introduce this tensor using the {\it TTC} function \verb+InputSRiemann+. This is internally equivalent to use \verb+InputTensor+ and \verb+InputSymmetries+ to declare the Riemann and Ricci tensors and to give the relation between them and between the scalar curvature.
The enumeration of the various Riemann rules used in the session comes from \cite{MathTensor}. 

On running the session we have used a Pentium-133 with 32Mb of RAM memory and the {\it Mathematica} version 2.2. 
The time to run the whole  session is about 1500 seconds. The maximum memory used is about 2.6 Mb (the memory used to store the whole set of rules for future calculus is about 1Mb). The {\it TTC} program takes about 1.7Mb of memory so the total memory used  by {\it Mathematica} is about 4.3Mb. 

\newpage

\begin{verbatim}

(*We have used  the RiemannRule## list from the MathTensor book. 
In this way these is a good test*)

In[]:=<<ttc/ttc.m

 _____________________________________________
         Tools of Tensor Calculus 4.0         
    by A.Balfagon, P.Castellvi and X.Jaen     
       http://baldufa.upc.es/ttc              
       e-mail: ttc@baldufa.upc.es             
______________________________________________
                          
TTCSettings=               
Compact[]=Off
SimplifyLevel=1
TTCSimplify={SinTosin, Together, sinToSin}
ScalarBasis=XX
ScalarBasisQ=True
Dimension[XX]=3
$RecursionLimit=5000
$SChristoffelTensor=G
______________________________________________

In[]:= InputSRiemann[gn,XX,"R",Rie,Ric,R]

{Rie, Ric, R}

In[]:= InputIndex[{i,j,k,l,m,n}]

{i, j, k, l, m, n}

(*RiemannRule1*)
In[]:= RR1=Rie[-a,a,-c,-d]//Index[gn]
           k 
0     + R
 i j     k   i j 

In[]:= RR1//Index[gn,SimplifyAllIndex]
0

(*RiemannRule6*)

In[]:= RR6=Rie[-a,-b,-c,-d] Ric[a,b]//Index[gn]
         k l 
0     + R     R
 i j           k l i j 

In[]:= RR6//Index[gn,SimplifyAllIndex]
 0

(*RiemannRule7*)
In[]:= RR7=Rie[-a,-b,-c,-d] Rie[-e,a,b,c]+
1/2 Rie[-p,-d,-q,-r] Rie[-e,p,q,r]//Index[gn]
           k l m 
        R         R
         j         k i l m       k l m 
0     + ------------------- + R         R
 i j             2             j         k l m i 
In[]:= RR7//Index[gn,SimplifyAllIndex]
0

(*RiemannRule10*)
In[]:= RR10=Ric[-a,-b] Rie[-c,-d,-e,a] Rie[b,c,d,e]+
1/2 Ric[-p,-q] Rie[-r,-s,-t,p]Rie[q,t,r,s]//Index[gn]
                                   j k l m         i 
                            R     R         R
       j k l m         i     i j             l m k 
R     R         R         + -------------------------
 i j             k l m                  2

In[]:= (RR10//Index[gn,SimplifyAllIndex])//Timing
0


(*RiemannRule13: here we have detected an error. In the Mathtensor
  book the second term has 1/4 as a factor. The correct factor is
  1/2*)
In[]:= RR13=Rie[-a,-b,-c,-d] Rie[-e,a,-g,b] Rie[c,d,e,g]-
1/2 Rie[-p,-q,-r,-s] Rie[-t,-u,p,q]Rie[r,s,t,u]//Index[gn]
           k l m n     i   j    
R         R         R         - 
 i j k l             m   n       

           k l m n       i j 
R         R         R        
 i j k l             m n 
----------------------------
             2
            
In[]:= RR13//Index[gn,SimplifyAllIndex]
0


(*RiemannRule15*)
In[]:= RR15=Rie[-a,-b,-c,-d] Rie[-e,a,-g,c] Rie[b,g,d,e]-
Rie[-p,-q,-r,-s] Rie[-t,p,-u,r] Rie[q,t,s,u]+
1/4 Rie[-p,-q,-r,-s] Rie[-t,-u,p,q]Rie[r,s,t,u]//Index[gn]
             j m l n     i   k 
-(R         R         R        ) + 
   i j k l             m   n 
 
            k l m n       i j 
 R         R         R
  i j k l             m n 
 ----------------------------- + 
               4
 
            j m l n     i   k 
 R         R         R
  i j k l             n   m 

In[]:= RR15//Index[gn,SimplifyAllIndex]
0

(*RiemannRule16*)
In[]:= RR16=Ric[-a,-b,.;b]-1/2 R[.;-a] //Index[gn]
       .;i 
 i    R        i   .;j 
0   - ----- + R
        2        j 
In[]:= RR16//Index[gn,SimplifyAllIndex]
0

(*RiemannRule18*)
In[]:= RR18=Ric[-a,-b,.;-c,.;b]-
1/2 R[.;-a,.;-c]- Ric[-p,-a]Ric[-c,p]+Ric[-p,-q] Rie[-a,q,-c,p] //Index[gn]
        R
         .;i .;j       k                   .;k             l   k 
0     - --------- - R     R     + R             + R     R
 i j        2        j     k i     i k .;j         k l   i   j 

In[]:= RR18//Index[gn,SimplifyAllIndex]
0

(*RiemannRule19*)
In[]:= RR19=Ric[-a,-b,.;-c,.;a]-
1/2 R[.;-b,.;-c]-Ric[-p,-b] Ric[-c,p]+
	Ric[-p,-q] Rie[-b,q,-c,p]//Index[gn]
        R
         .;i .;j       k                   .;k             l   k 
0     - --------- - R     R     + R             + R     R
 i j        2        j     k i     k i .;j         k l   i   j 

In[]:= RR19//Index[gn,SimplifyAllIndex]
0


(*RiemannRule24*)
In[]:= RR24=R[.;-a,.;-b,.;a,.;b]-
1/2 R[.;-p] R[.;p]-
R[.;-p,.;p,.;-q,.;q]-
R[.;-p,.;-q] Ric[p,q]//Index[gn]
         .;i 
-(R     R    )
   .;i                .;i     .;j             .;i .;j 
-------------- - R                 + R                 - 
      2           .;i     .;j         .;i .;j 
 
            i j 
 R         R
  .;i .;j 
In[]:= RR24//Index[gn,SimplifyAllIndex]
0

\end{verbatim}

\end{document}